\documentclass[onecolumn,aps,amsmath,amssymb,amsfonts,floatfix]{revtex4}
\usepackage{graphicx}% Include figure files
\usepackage{dcolumn}% Align table columns on decimal point
\usepackage{bm}% bold math
\usepackage[hypertex]{hyperref}
\usepackage{latexsym}
\usepackage{float}
\usepackage{supertabular}
\usepackage{longtable}

\begin{document}
\title{Large Fluctuations and Fixation in Evolutionary Games}
\author{Michael Assaf$^1$ and Mauro Mobilia$^{2}$}
\affiliation{$^1$Racah Institute of Physics, Hebrew University of
Jerusalem, Jerusalem 91904, Israel\\$^2$Department of Applied Mathematics, University of Leeds, Leeds LS2 9JT, United Kingdom
\\E-mail: assaf@phys.huji.ac.il, m.mobilia@leeds.ac.uk}
\begin{abstract}
\large\textbf{Abstract.} \normalsize
We study large fluctuations in evolutionary games belonging to the coordination and anti-coordination classes. The dynamics of these games, modeling cooperation dilemmas, is characterized by a coexistence fixed point separating two absorbing states. We are particularly interested in the problem of fixation that refers to the possibility that a few mutants take over the entire population. Here, the fixation phenomenon is induced by large fluctuations and is investigated by a semi-classical WKB (Wentzel-Kramers-Brillouin) theory generalized to treat stochastic systems possessing multiple absorbing states. Importantly, this method allows us to analyze the combined influence of selection and random fluctuations on the evolutionary dynamics \textit{beyond} the weak selection limit often considered in previous works. We accurately compute, including pre-exponential factors, the probability distribution function in the long-lived coexistence state and the mean fixation time necessary for a few mutants to take over the entire population in anti-coordination games, and also the fixation probability in the coordination class. Our analytical results compare excellently with extensive numerical simulations. Furthermore, we demonstrate that our treatment is superior to the Fokker-Planck approximation when the selection intensity is finite.
\end{abstract}
\maketitle

\tableofcontents

\section{Introduction \& Models}
Systems in which successful strategies spread by imitation or reproduction can be
described by evolutionary game theory (EGT), whose prototypical models are commonly studied in the context of evolutionary biology, ecology, sociology and economics~\cite{Maynard,Hofbauer,Nowak,Szabo,Hamilton,Gintis}. Recently, it has also
been realized that techniques of statistical physics can help gain further insight into this interdisciplinary area~\cite{Szabo,Szabo2002}.
Originally, EGT was formulated in terms of deterministic replicator equations
valid to treat populations of infinite size~\cite{Maynard,Hofbauer,Szabo} and related to
the  celebrated Lotka-Volterra equations~\cite{LV,May74,Hofbauer,Nowak,Szabo,Mobilia-2007}.
However, it has been long recognized that the picture emerging from replicator dynamics is often fundamentally altered
by demographic fluctuations.

One of the most striking effects of fluctuations in EGT is {\it fixation} that refers
to  the possibility for a few ``mutants''  to take over (fixate) an entire population causing the extinction of the ``wild species''.
To rationalize the effect of stochasticity in finite populations, Nowak {\it et al.}~\cite{NowakTaylor,Nowak} introduced a parameter $w$ controlling  the interplay between random (demographic) fluctuations  and selection (leading to nonlinear effects). This approach reflects the fact that evolutionary processes comprise two main competing mechanisms.
On the one hand, there is selection by individuals' different fitness (reproductive potential), which underlies adaptation~\cite{PopGen,Maynard,Hofbauer,Nowak,Trivers,Axelrod,Hamilton,Nowak2005}. On the other hand, in all finite-size populations birth and death events cause random (demographic) fluctuations, which play a central role in neutral theories where selection is considered of marginal importance~\cite{Kimura,Hubbell,Bell,Blythe}.

Most of the analytical  results concerning EGT in finite population have been obtained in the \textit{weak selection limit} ($w \to 0$),
which  is often biologically relevant~\cite{Ohta} and greatly simplifies the mathematical analysis.
In particular, the fixation probability of a species under frequency-dependent selection in a finite population (of two species) has been computed in this limit~\cite{NowakTaylor}.  However, very different behaviors have been obtained under strong and weak selection~(see, e.g., \cite{Antal,Traulsen2006,Altrock,Santos,Ohtsuki}) and  the respective influence  of selection and demographic fluctuations on evolution still remains to be fully understood. Our purpose in this paper is to study fixation resulting from large fluctuations in two classes of evolutionary games (the anti-coordination and coordination classes, see below) under arbitrary (finite) selection strength, and to elucidate the nontrivial interplay between selection intensity and demographic fluctuations.

For the sake concreteness, throughout this paper we will consider (symmetric) $2 \times 2$ evolutionary games. Here, one has a homogeneous (well-mixed) population comprising a total of $N$ individuals, $n$ are of the ``mutant'' species $\textsf{A}$ and $N-n$
are of the ``wild'' type $\textsf{B}$.
As usual, it is assumed that there are pairwise and symmetric interactions between individuals drawn at random. The reproductive potential of an individual is specified by the payoff of interaction with others \cite{Hofbauer,Szabo,Nowak,Maynard}. Specifically, when two individuals of species
$\textsf{A}$ interact, both receive a payoff $a$. The interaction of a pair of individuals of different species yields
 payoffs $b$ and $c$ to the $\textsf{A}$ and $\textsf{B}$ individuals,  respectively.
Similarly, when two
individuals of type
$\textsf{B}$ interact, both get a payoff $d$. Therefore, the respective average (per individual) payoffs of species $\textsf{A}$ and $\textsf{B}$ read:
$\Pi_{A}(n)=(n-1)/(N-1)a + (N-n)/(N-1)b$, and $\Pi_{B}(n)=n/(N-1)c + (N-n-1)/(N-1)d$, with self-interactions being excluded. It is useful to introduce the difference between the average payoffs $\Delta \Pi(n)=\Pi_{A}(n) - \Pi_{B}(n)= (a-b-c+d)n/(N-1) - (d-b)N/(N-1)-(a-d)/(N-1)$. For populations of infinite size ($N \to \infty$), the dynamics is of mean-field type and   commonly described by the replicator equations. The latter are obtained from the payoff matrix  by comparing the success of  a given type with  the population average~\cite{Hofbauer,Nowak,Szabo}. In this limit $x\equiv n/N$ can be regarded as a continuous variable and the dynamics is specified by the following replicator equation:
\begin{eqnarray}
\label{Repl}
\dot{x}=x(1-x)\Delta \Pi(x)\,,\;\;\;\Delta\Pi(x)=(a-c) x - (d-b)(1-x).
\end{eqnarray}
This equation is characterized by two absorbing fixed points $x=x_A^*=1$ and $x=x_B^*=0$ corresponding to a system with all $\textsf{A}$'s and $\textsf{B}$'s, respectively. Moreover, there can be an interior fixed point obtained by solving $\Delta \Pi(x^*)=0$
\begin{eqnarray}
\label{IntFP}
x^*=\frac{d-b}{a-b-c+d}.
\end{eqnarray}
Depending on the entries of the payoff matrix, one has various classes of games
representing models of cooperation dilemmas~\cite{Nowak2005,Szabo,Traulsen2006}.
When one species always dominates the other, one has the dominance class, where $\textsf{A}$
is the dominant type when $a>c$ and $b>d$, while $\textsf{B}$ is dominant when $a<c$ and $b<d$. In this work we are interested in the other two classes, anti-coordination games (ACG) and coordination games (CG), where there is an interior fixed point corresponding to a coexistence state. In the class of ACG (\textit{e.g.}, ``snowdrift'' and ``hawk-dove'' games~\cite{Maynard,Hofbauer,Szabo}),
$c>a$, $b>d$, and $x^*$ is an attractor corresponding to the stable coexistence of $\textsf{A}$ and
$\textsf{B}$ types. Here, the absorbing states $x_A^*$ and $x_B^*$ are (evolutionary) unstable. On the other hand, in the class of CG
 (\textit{e.g.} ``stag-hunt" game~\cite{Hofbauer,Nowak,Szabo}) $a>c$, $d>b$, and
$x^*$ is repelling, while the absorbing states are (evolutionary) stable.

Fluctuations arising from the discreteness of individuals and from the stochastic nature of the interactions may drastically alter the predictions of the deterministic replicator equations~\cite{noise}. In particular, a few mutants can always attain {\it fixation} by taking over  the entire population (see below). The resulting \textit{stochastic} evolutionary dynamics is aptly described in terms of single-step birth-death processes~\cite{Gardiner,vanK,Ewens},
for which a key quantity is the   time-dependent probability distribution function (PDF) $P_n (t)$ of population sizes.
The latter gives  the probability of finding the system in a state with $n$ individuals of species  $\textsf{A}$ at time $t$, and obeys the following master equation:
\begin{eqnarray}\label{master}
\frac{d P_n(t)}{dt} = T^+(n-1) P_{n-1}+T^-(n+1) P_{n+1}-[T^+(n)+T^-(n)]P_n.
\end{eqnarray}
Here $T^+(n)$ and $T^-(n)$ respectively denote the transition rates from a state with $n$ $\,\textsf{A}$'s to a state with $n+1$ and $n-1$ $\,\textsf{A}$'s.  As the state space is bounded, $n\in[0,N]$, and $n=0$ and $n=N$ are absorbing states, the transition rates at the boundaries satisfy $T^{\pm}(0)=T^{\pm}(N)=0$.

According to general prescriptions of EGT, the transition rates are functions of each species' fitness (effective potential to reproduce),
$f_{\sigma}$, with $\sigma\in (\textsf{A},\textsf{B})$, \textit{i.e.}  $T^{\pm}(n)=T^{\pm}[f_{\sigma}(n)]$. The fitness of an individual of species $\sigma$ reads~\cite{NowakTaylor,Nowak} $\,f_{\sigma}(n)\equiv 1-w+w\Pi_{\sigma}(n)$. Here, the interplay between random fluctuations and selection is controlled by the parameter $w$ (with $0 \leq  w \leq 1$), where in the neutral case, $w=0$, there is no selection (but only random fluctuations), while in the strong selection regime, $w=1$, the influence of random fluctuations is negligible.

In this paper we consider the following update rules commonly used in the EGT literature, specifying the functional dependence of $T^{\pm}$ on the species fitnesses.
For the frequency-dependent Moran Process (fMP)~\cite{Nowak,Szabo,Hauert-preprint}, one has
\begin{eqnarray}
\label{MP}
T^{+}(n) = \frac{f_A(n)}{\bar{f}(n)}\,\Phi(n)\;,\;\;\;\;T^{-}(n)= \frac{f_B(n)}{\bar{f}(n)}\,\Phi(n),
\end{eqnarray}
where $\bar{f}(n)=[n f_A(n)  + (N-n)f_B(n)]/N$ is the population average fitness and $\Phi(n)\equiv n(N-n)/N^2.$
For the linear Moran process (LMP)~\cite{Traulsen,Hauert-preprint}, the transition rates read
\begin{eqnarray}
\label{LMP}
T^{+}(n) = \frac{1}{2}\{1+ [f_A(n)- \bar{f}(n)]\}\,\Phi(n)\;,\;\;\;\;T^{-}(n)= \frac{1}{2}\{1+ [f_B(n)- \bar{f}(n)]\}\,\Phi(n).
\end{eqnarray}
As the LMP is obtained from a small $w$ expansion of the fMP, (\ref{LMP}) can be regarded as the ``weak selection'' counterpart of the rates~(\ref{MP}).
A process closely related to the LMP is the ``local update'' process (LUP)~\cite{Traulsen,Hauert-preprint} with rates:
\begin{eqnarray}
\label{LU}
T^{+}(n)= \frac{1}{2}\{1+ [f_A(n)- f_B(n)]\}\,\Phi(n)\;,\;\;\;\;T^{-}(n)= \frac{1}{2}\{1+ [f_B(n)- f_A(n)]\}\,\Phi(n).
\end{eqnarray}
Here, for simplicity and without restriction, we have assumed that the maximum payoff difference is $1$~\cite{Hauert-preprint}.
Finally, for the Fermi process (FP)~\cite{Szabo2002,Traulsen2006,Altrock,Hauert-preprint}, one has
\begin{eqnarray}
\label{FP}
T^{+}(n) = \left\{1+{\rm exp} [f_B(n)- f_A(n)]\right\}^{-1}  \Phi(n)\;,\;\;\;\;T^{-}(n)&=& \left\{1+{\rm exp} [f_A(n)- f_B(n)]\right\}^{-1} \Phi(n).
\end{eqnarray}
In the following, we omit in all these cases the self-interaction terms in the expressions of $\Pi_A (n)$ and $\Pi_B (n)$~\cite{o1}.
Note, that multiplying both sides of (\ref{master}) by $n$ and summing over all $n$'s, one obtains an equation for the mean
number of $\textsf{A}$ individuals. In the leading order of $N\gg 1$ and upon rescaling time, one arrives at the following rate equation for
 the concentration of mutants: $\dot{x}=T^+(Nx)-T^-(Nx)$.
Such a replicator-like equation  shares the same properties (fixed points and stability) of Eq.~(\ref{Repl}), but generally differs from it when $\bar{f}$ is
non-constant~(see e.g., \cite{Maynard,Traulsen}).

Let us denote by $\phi_i^A$ the  probability that $i$  mutants of species $\textsf{A}$ (usually $i\ll N$) replace all the individuals of the wild type $\textsf{B}$. That is, $\phi_i^A$ is the {\it probability of fixation} of the $\textsf{A}$ species starting with $i$ mutants. The {\it conditional and unconditional mean fixation times} (MFTs) $\tau_{i}^{A}$ and $\tau_{i}$, respectively, are the times associated with the fixation event. The former, $\tau_{i}^{A}$, is the average time it takes for $i$ mutants of species $\textsf{A}$ to take over the population, while the latter, $\tau_{i}$, is the mean time it takes the population, initially comprising $i$ individuals of species $\textsf{A}$, to become homogeneous again (\textit{i.e.} populated either by all  $\textsf{A}$'s or all $\textsf{B}$'s). For all one-dimensional single-step birth-death processes, as those defined by (\ref{MP}-\ref{FP}), there are exact formulas for the above quantities~\cite{vanK,Gardiner,Antal,Altrock}. For instance, the fixation probability reads
\begin{eqnarray}
\label{ProbFix}
\phi_i^{A}=\frac{1+\sum_{k=1}^{i-1}\prod_{l=1}^{k} \gamma_l}{1+\sum_{k=1}^{N-1}\prod_{l=1}^{k} \gamma_l},
\end{eqnarray}
where  $\gamma_i=T^{-}(i)/T^{+}(i)$, while for the unconditional MFT, one has
\begin{eqnarray}
\label{ProbFix1}
\tau_i= -\tau_1 \sum_{k=i}^{N-1} \prod_{m=1}^{k}\gamma_m +  \sum_{k=i}^{N-1}\sum_{l=1}^{k} \frac{1}{T^+ (l)} \prod_{m=l+1}^{k} \gamma_m\;;\;\;\;\;\;\;\tau_1=\phi_1^{A} \sum_{k=1}^{N-1}\sum_{l=1}^{k} \frac{1}{T^+ (l)} \prod_{m=l+1}^{k} \gamma_m.
\end{eqnarray}
Even though the expressions (\ref{ProbFix}-\ref{ProbFix1}) are exact, they are unwieldy and cannot be  generalized to multi-step processes and to $s \times s$ games, with $s>2$. Furthermore, it is highly nontrivial to extract their asymptotic behavior. In fact, with the exception of Ref.~\cite{Antal} where
the fixation probability and  MFTs  were calculated in the leading order for the fMP (focusing on $w=1$), most of the analytical results in the literature have been obtained in the weak selection limit, $Nw \ll 1$, often using the Fokker-Planck equation (FPE)~\cite{Pacheco,Traulsen2006,Altrock,Hauert-preprint}.

In this paper we go beyond the weak selection limit and investigate fixation phenomena induced by large fluctuations in the classes of ACG and CG. Our approach relies on the WKB approximation~\cite{Bender} applied directly to the master equation~(\ref{master})~\cite{kubo,dykman,EsK,AM}, that we generalize to account for the existence of multiple absorbing states. Here, the WKB approximation is an asymptotic series expansion in powers of $1/N \ll 1$ based on an exponential ansatz~\cite{analogy} [see Eq.~(\ref{ansatz})] and differs from the van-Kampen system size expansion that yields the FPE~\cite{Gardiner,vanK} (see Sec. III.B). With the WKB approach and for any finite selection intensity, we accurately compute the MFTs and fixation probability (including pre-exponential factors) for generic transition rates. Our general results are then applied to the processes defined by the transition rates~(\ref{MP}-\ref{FP}) and successfully compared with extensive numerical simulations. The predictions of our WKB approach are also shown to be advantageous over those of the FPE when $w$ is finite (see also Refs.~\cite{fpill}).

The remainder of this paper, of which a brief account has recently been given in Ref.~\cite{MA}, is organized as follows. The next section is dedicated to the ACG, for which a general treatment is presented in Section II.A, while applications are discussed in Section II.B. Section III concerns the CG class, with general results and applications discussed in Sections III.A and III.B, respectively. Finally, we give a summary of our findings and present our conclusions in Section IV. Some technical details are relegated in an appendix.

\section{Anti-coordination games: Metastability, fixation times and probability}
In this section we use the WKB approach to investigate large fluctuations in systems of evolutionary games characterized by {\it metastability}.
For $N\gg 1$, in general this case arises in the ACG, to which  the snowdrift and hawk-dove games belong~\cite{Hofbauer,Nowak,Szabo}. In ACG, the elements of the payoff matrix  satisfy $b>d$ and $c>a$. Here, the attracting (interior) fixed point $x^*$ [see Eq.~(\ref{IntFP})] in the language of the replicator equation~(\ref{Repl}), separates the repelling fixed points $x_A^*$ (all $\textsf{A}$'s) and $x_B^*$ (all $\textsf{B}$'s).
In the presence of internal noise, $x^*$ corresponds to a long-lived \textit{metastable} state, where after a sufficiently long time the system
is eventually driven into one of the two absorbing states via a large fluctuation.
In this section we first derive general results concerning the metastable dynamics of stochastic systems possessing two absorbing states. Then, using the transition rates (\ref{MP})-(\ref{FP}), we apply these findings to the case of ACG.

\subsection{General treatment \& results}
Our starting point is the master equation~(\ref{master}). In case of a finite space $n\in[0,N]$, one can always expand $P_n(t)$ in a finite series of eigenvectors and eigenvalues of the stochastic generator associated with the Markov chain (\ref{master}). We assume here and henceforth that $N\gg 1$. In this case, for any (sufficiently large) given initial population of $\textsf{A}$'s, after a relaxation time scale $t_r$, the system converges into a long-lived metastable (coexistence) state \textit{prior} to fixation of either species. This metastable state corresponds to a PDF of population sizes peaked in the vicinity of $\bar{n}_*=Nx^*$, where $x^*$ is attracting interior fixed point of (\ref{Repl}). Here, the MFT $\tau$ is associated with the slow decay of the metastable state, see below, and satisfies $t_r\ll \tau$. That is, fixation  occurs  in the aftermath of a long-lasting coexistence state.

It turns out, that for times $t\gg t_r$, when the system has already converged into the metastable state, the higher excited eigenvectors have already decayed (see, \textit{e.g.}, Ref.~\cite{Assaf}). At such times, only the first excited eigenvector of (\ref{master}), $\pi_n$, called the quasi-stationary distribution (QSD), survives and determines the \textit{shape} of the metastable PDF. Correspondingly, the decay rate of the metastable PDF is determined by the first-excited eigenvalue of~(\ref{master})~\cite{Assaf}. While the metastable PDF decays, the probabilities $P_0(t)$ and $P_N(t)$ slowly grow  in time and, at $t\to\infty$,
reach some final values such that $P_0(\infty)+P_N(\infty)=1$. Therefore, at $t\gg t_r$, one can write
\begin{eqnarray}\label{qsd}
P_{n=1,\dots,N-1}(t)\simeq \pi_n e^{-t/\tau}\;,\;\;\;P_0(t)\simeq \phi (1-e^{-t/\tau})\;,\;\;\;P_N(t)\simeq (1-\phi)(1-e^{-t/\tau}).
\end{eqnarray}
From this metastable dynamics one can immediately see that $\tau$ is the (unconditional) MFT, while $\phi=\phi^B$ is the fixation probability of species $\textsf{B}$, and $1-\phi$ is the fixation probability of species $\textsf{A}$.

It follows from (\ref{qsd}) that $\dot{P}_0+\dot{P}_N=(1/\tau)e^{-t/\tau}$, while  from the master equation (\ref{master}) one has $\dot{P}_N=T^+(N-1)\pi_{N-1}e^{-t/\tau}$ and $\dot{P}_0=T^-(1)\pi_{1}e^{-t/\tau}$. We thus obtain
$\dot{P}_0(t)=\left[\tau^{-1}-T^+(N-1)\pi_{N-1}\right]e^{-t/\tau}=T^-(1)\pi_{1}e^{-t/\tau}$, whose solution (with $P_0(0)=0$) is
$P_0(t)=\left[1-\tau T^+(N-1)\pi_{N-1}\right](1-e^{-t/\tau})$. Using this solution and Eq.~(\ref{qsd}), we
obtain the fixation probability $\phi$ which turns out to be the relative flux into the absorbing state $n=0$. Moreover, as the unconditional MFT is the (inverse of the) decay rate of the metastable state, it is given by the (inverse of the) sum of probability fluxes into the two absorbing states . Thus, we have
\begin{equation}\label{mtephi}
\phi=T^-(1)\pi_{1}\tau \;,\;\;\tau=[T^-(1)\pi_{1}+T^+(N-1)\pi_{N-1}]^{-1},
\end{equation}
where the unknowns $\pi_1$ and $\pi_{N-1}$  will be determined shortly. Correspondingly, $\tau^A$ and $\tau^B$ -- the \textit{conditional MFTs}  of
species $\textsf{A}$ and $\textsf{B}$ respectively, can also be found. The former is the mean time it takes the $\textsf{A}$ species to fixate conditioned on the non-fixation of the $\textsf{B}$ species;
it is determined by the the inverse of the flux to the state $n=N$.
Using the same reasoning for $\tau^B$, we thus have $\tau^A=[T^+(N-1)\pi_{N-1}]^{-1}$ and $\tau^B=[T^-(1)\pi_{1}]^{-1}$.
Note, that since we have assumed that the system reaches the metastable state prior to fixation, our results~(\ref{mtephi}) are independent of the initial condition. As we shall see below, this assumption holds for ACG when the selection strength is finite.

Substituting the metastable ansatz~(\ref{qsd}) into Eq.~(\ref{master}), we arrive at the quasi-stationary master equation (QSME). Neglecting the
exponentially small term $\pi_n/\tau$ (to be verified a-posteriori) on the left-hand-side, the QSME becomes
\begin{eqnarray}\label{masterqsd}
0 = T^+(n-1) \pi_{n-1}+T^-(n+1) \pi_{n+1}-[T^+(n)+T^-(n)]\pi_n.
\end{eqnarray}

In the following, this equation is analyzed by using the WKB approximation. Our aim, in addition  to finding the fixation probability and MFTs, is to find the complete QSD, $\pi_n$, and demonstrate the non-Gaussian nature of its tails. To do so, and since there are two absorbing states in this problem, we solve the QSME~(\ref{masterqsd}) separately in \textit{three} overlapping regions: in the bulk and not too close to the absorbing boundaries, and in the close vicinities of $n=0$ and $n=N$. These solutions are then matched in their regions of joint validity.

In the bulk region (whose accurate boundaries are specified below) we employ the WKB ansatz
\begin{equation}
\pi_n\equiv \pi_{xN}=\pi(x)={\cal A}e^{-NS(x)-S_1(x)}\,, \label{ansatz}
\end{equation}
where $N\gg 1$, and we have introduced the rescaled coordinate $x=n/N$.
Here, $S(x)$ is the action while $S_1(x)$ is the amplitude.
To have a consistent perturbation theory, we assume that these quantities are smooth functions of order unity. The constant prefactor ${\cal A}$ is introduced for
technical convenience. It is convenient to rewrite the transition rates as continuous functions
${\cal T}_{\pm}(x)\equiv T^{\pm}(n)$ of the rescaled continuous coordinate $0\leq x\leq 1$. We will assume that in the bulk, ${\cal T}_{\pm}(x)={\cal O}(1)$, which is  satisfied by all the transition rates (\ref{MP})-(\ref{FP}).

Plugging ansatz~(\ref{ansatz}) into Eq.~(\ref{masterqsd}), and expanding the functions of $x\pm N^{-1}$ up to ${\cal O}(N^{-1})$, we arrive at
\begin{eqnarray}
\pi(x)\!\left\{{\cal T}_+(x)\left[e^{S'}\!\left(1-\frac{S''}{2N}+\frac{S_1'}{N}\right)-1\right]+{\cal T}_-(x)\left[e^{-S'}\!\left(1-\frac{S''}{2N}-\frac{S_1'}{N}\right)-1\right]+\frac{1}{N}\left[e^{-S'}{\cal T}_-'(x)-e^{S'}{\cal T}_+'(x)\right]\!\right\}=0.
\end{eqnarray}
This equation can be solved order by order in $N\gg 1$. In the leading ${\cal O}(1)$ order, one obtains a stationary Hamilton-Jacobi equation for the action $S(x)$, $H[x,S'(x)]=0$ with a Hamiltonian given by
\begin{equation}\label{singleHam}
H(x,p)={\cal T}_+(x)(e^p-1)+{\cal T}_{-}(x)(e^{-p}-1)\,,
\end{equation}
where we have defined the auxiliary momentum coordinate $p(x)=S'(x)$~\cite{dykman}. Therefore, the leading-order calculations correspond to finding a nontrivial zero-energy trajectory of the auxiliary Hamiltonian~(\ref{singleHam})~\cite{dykman}. It turns out that there is exactly one (real) such trajectory~\cite{AM}, $p_a(x)$, called the activation trajectory. It represents the ``optimal-path'' along which the stochastic system evolves, almost with certainty, from the metastable state towards fixation. Here, the solution of  $H[x,p_a(x)]=0$ is $p_a(x)=-\ln [{\cal T}_{+}(x)/{\cal T}_-(x)]$~\cite{AM}. The corresponding action along this trajectory is
\begin{equation}\label{Ssingle}
S(x)=-\int^x\ln [{\cal T}_{+}(\xi)/{\cal T}_-(\xi)]\,d\xi.
\end{equation}
In the subleading ${\cal O}(1/N)$ order, one arrives at a first-order transport-like differential equation for $S_1(x)$, whose solution is~\cite{EsK,AM}
\begin{eqnarray}\label{S11}
\hspace{-2mm}S_1(x)=\frac{1}{2}\ln [{\cal T}_+(x) {\cal T}_-(x)].
\end{eqnarray}
Therefore, the solution in the bulk region can be written as
\begin{eqnarray}\label{fastmode0}
\pi(x) =\frac{{\cal A}}{\sqrt{{\cal T}_+(x){\cal T}_-(x)}}e^{N\int^x\ln  [{\cal T}_{+}(\xi)/{\cal T}_-(\xi)]d\xi}\,.
\end{eqnarray}
It is worth emphasizing that this solution is valid in the regime where ${\cal T}_{\pm}(x)={\cal O}(1)$, \textit{i.e.}, not too close to the absorbing boundaries $x=0$ and $x=1$, where the transition rates vanish. As for $N \gg 1$ the QSD is strongly peaked in the vicinity of the attracting fixed point $x^*$, the constant ${\cal A}$ can be determined by normalizing to unity the Gaussian asymptote of the QSD around $x^*$. The latter is obtained by expanding Eq.~(\ref{ansatz}) to second order about $x^*$  and using $p_a(x^*)=0$ [since ${\cal T}_+(x^*)={\cal T}_-(x^*)$]. Normalizing the resulting Gaussian asymptote,
$\pi(x)\simeq {\cal A} e^{-NS(x^*)-S_1(x^*)-(N/2)S^{\prime\prime}(x^*)(x-x^*)^2}$, yields the constant ${\cal A}$, and therefore the QSD is given by
\begin{eqnarray}\label{fastmode}
\pi(x) ={\cal T}_+(x^*)\,\sqrt{\frac{S''(x^*)}{2\pi N\,{\cal T}_+(x){\cal T}_-(x)}}\,e^{-N[S(x)-S(x^*)]}\,.
\end{eqnarray}
Here, $S''(x^*)={\cal T}_-^{\prime}(x^*)/{\cal T}_-(x^*)-{\cal T}_+^{\prime}(x^*)/{\cal T}_+(x^*)>0$, as $x^*$ is an attracting fixed point and so ${\cal T}_+'(x^*)-{\cal T}_-'(x^*)<0$.

We now turn to dealing with the QSME~(\ref{masterqsd}) in the close vicinities of the absorbing boundaries where the WKB approximation breaks down. First, expanding the transition rates ${\cal T}_{\pm}(x)\simeq x {\cal T}_{\pm}'(0)$ about  $x=0$ [where ${\cal T}_{\pm}(0)=0$], and multiplying Eq.~(\ref{masterqsd}) by $N$, one has
$\,0={\cal T}_+^{\prime}(0)(n-1)\pi_{n-1}+{\cal T}_-^{\prime}(0)(n+1)\pi_{n+1}-n[{\cal T}_+^{\prime}(0)+{\cal T}_-^{\prime}(0)]\pi_n$,
whose recursive solution is \cite{AM}
\begin{equation}\label{rec1}
\pi_n=\frac{\left(R_0^n-1\right)\pi_1}{\left(R_0-1\right)n}.
\end{equation}
Here we have introduced the parameter $R_0\equiv {\cal T}_+^{\prime}(0)/{\cal T}_-^{\prime}(0)$.
This procedure turns out to be valid in the range $1\leq n\ll \sqrt{N}$~\cite{AM}.
Similarly, close to the boundary  $n=N$, the rates in the QSME (\ref{masterqsd}) can be expanded in the vicinity of $x=1$ yielding $\,0={\cal T}_+^{\prime}(1)(N-n+1)\pi_{n-1}+{\cal T}_-^{\prime}(1)(N-n-1)\pi_{n+1}-(N-n)[{\cal T}_+^{\prime}(1)+{\cal T}_-^{\prime}(1)]\pi_n$. The solution of this equation, valid for  $1\leq N-n\ll \sqrt{N}$ and with $R_1\equiv {\cal T}_-^{\prime}(1)/{\cal T}_+^{\prime}(1)$, satisfies
\begin{equation}\label{recn}
\pi_n=\frac{\left(R_1^{N-n}-1\right)\pi_{N-1}}{\left(R_1-1\right)(N-n)}\,.
\end{equation}

We are in the position to find the complete QSD by matching Eq.~(\ref{rec1}) and (\ref{recn}) with the asymptotes of (\ref{fastmode}) in the regions $1\ll n\ll \sqrt{N}$ and $1\ll N-n\ll \sqrt{N}$, respectively. In the vicinity of $x=0$, the asymptote of (\ref{fastmode}) can be found by writing $S(x)\simeq S(0)+x p_a(0)$, with $p_a(0)=\ln [{\cal T}_-^{\prime}(0)/{\cal T}_+^{\prime}(0)]$, which yields
\begin{eqnarray}\label{fastmodeasym0}
\pi(x) =\frac{{\cal T}_+(x^*)\sqrt{S''(x^*)}}{x\sqrt{2\pi N\,{\cal T}_+^{\prime}(0){\cal T}_-^{\prime}(0)}}\,R_0^{Nx}
e^{-N[S(0)-S(x^*)]}.
\end{eqnarray}
This asymptote is valid for $1\ll n\ll \sqrt{N}$~\cite{AM}. Similarly, the asymptote of Eq.~(\ref{fastmode}) in the vicinity of $x=1$ reads
\begin{eqnarray}\label{fastmodeasymn}
\pi(x) =\frac{{\cal T}_+(x^*)\sqrt{S''(x^*)} R_1^{N(1-x)}}{(1-x)\sqrt{2\pi N\,{\cal T}_+^{\prime}(1){\cal T}_-^{\prime}(1)}}\, e^{-N[S(1)-S(x^*)]},
\end{eqnarray}
and is valid for $1\ll N-n\ll \sqrt{N}$. Matching Eqs.~(\ref{fastmodeasym0}) and (\ref{fastmodeasymn}), respectively with Eqs.~(\ref{rec1}) and (\ref{recn}) yields
\begin{eqnarray}\label{pi1n}
\pi_1=\sqrt{\frac{N S''(x^*)}{2\pi}}\frac{{\cal T}_+(x^*)\left(R_0-1\right)}{\sqrt{{\cal T}_+^{\prime}(0){\cal T}_-^{\prime}(0)}}e^{-N[S(0)-S(x^*)]}\;,\;\;\;\;\pi_{N-1}=\sqrt{\frac{N S''(x^*)}{2\pi}}\frac{{\cal T}_+(x^*)\left(R_1-1\right)}{\sqrt{{\cal T}_+^{\prime}(1){\cal T}_-^{\prime}(1)}}e^{-N[S(1)-S(x^*)]}.
\end{eqnarray}

With the expressions (\ref{pi1n}) and (\ref{fastmode}), the QSD has been completely determined. The fixation probability and the MFTs can then be computed  according to (\ref{mtephi}). In fact, as $T^-(1)\simeq (1/N){\cal T}_-'(0)$ and $T^+(N-1)\simeq (1/N)|{\cal T}_+'(1)|$ [as ${\cal T}_+'(1)<0$], the fixation probability and unconditional MFT~(\ref{mtephi}) become
\begin{equation}\label{phimte1}
\phi=\frac{{\cal T}_-'(0)\pi_{1}}{{\cal T}_-'(0)\pi_{1}+|{\cal T}_+'(1)|\pi_{N-1}}\;,\;\;\;\;\tau=\frac{N}{{\cal T}_-'(0)\pi_1+|{\cal T}_+'(1)| \pi_{N-1}},
\end{equation}
while the conditional MFTs of species $\textsf{A}$ and  $\textsf{B}$ are respectively $\tau^A=N[|{\cal T}_+'(1)|\pi_{N-1}]^{-1}$ and $\tau^B=N[{\cal T}_-'(0)\pi_1]^{-1}$.
Importantly, since we have assumed that $\tau$ is exponentially large in $N$, these results indicate that our theory is valid as long as $N[S(1)-S(x^*)]\gg 1$ and $N[S(0)-S(x^*)]\gg 1$.

\subsection{Applications}
As applications of the general results that we have derived, we now explicitly consider the four types of update rules mentioned above, \textit{i.e.}, the fMP,
LMP, LUP and the FP (\ref{MP})-(\ref{FP}). For each of them we obtain the QSD, the MFTs and the fixation
probability under arbitrary (but finite) selection strength $0<w\leq 1$.
\begin{figure}
\includegraphics[width=3.2in, height=2.35in,clip=]{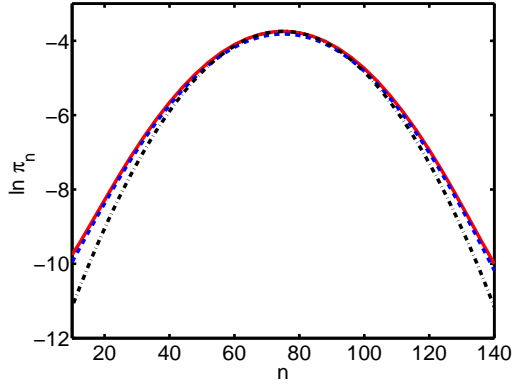}
\caption{{\it (Color online)}. Shown is $\ln \pi_n$ as a function of $n$
for $a=0.1$, $b=0.7$, $c=0.6$, $d=0.2$, $w=0.4$ and $N=150$, so that $n_*=75$. The dynamics is
implemented according to the fMP (\ref{MP}).
We compare the analytical result (\ref{qsdMP}) (solid line) with the numerical solution of the master equation (\ref{master}) (dashed line) with rates (\ref{MPrates}), and with a Gaussian approximation of this distribution with mean $n_*$ and standard deviation $\sigma=\sqrt{N/S''(x^*)}$ (dash-dotted line).
Near the tails, one can clearly observe the non-Gaussian nature of the QSD. Note that, very close to the boundaries at $n={\cal O}(1)$ and $N-n={\cal O}(1)$ (see text), Eq.~(\ref{qsdMP}) has to be replaced by  Eqs.~(\ref{rec1}) and (\ref{recn}).} \label{qsdmp}
\end{figure}
\subsubsection{Frequency-dependent Moran process}
For the fMP, the birth and death rates (\ref{MP}) in terms of the variable $x$ become
\begin{eqnarray}\label{MPrates}
{\cal T}_+(x)=\frac{\left\{1-w+w[a x +b(1-x)]\right\}x(1-x)}{1-w+w\left[ax^2+(b+c)x(1-x)+d(1-x)^2\right]}\,,\;\;
{\cal T}_-(x)=\frac{\left\{1-w+w[c x +d(1-x)]\right\}x(1-x)}{1-w+w\left[ax^2+(b+c)x(1-x)+d(1-x)^2\right]}
\end{eqnarray}
and one can check that ${\cal T}_-'(0)=|{\cal T}_+'(1)|=1$. The action $S(x)$  is computed from Eq.~(\ref{Ssingle})
with the rates (\ref{MPrates}), yielding
\begin{equation}\label{MPact}
S(x)=\int^x \ln\left\{\frac{1-w+w[c q +d(1-q)]}{1-w+w[a q +b(1-q)]}\right\}dq\,.
\end{equation}
For further analytical treatment, it is convenient to introduce the following parameters (also used in Ref.~\cite{Claussen}):
$A=1-w+wa$, $B=1-w+wb$, $C=1-w+wc$, and $D=1-w+wd$, where for ACG, $C>A$ and $B>D$. Performing the integral~(\ref{MPact}), one obtains after some algebra
\begin{eqnarray}\label{MPS}
e^{-NS(x)}=\left[Ax+B(1-x)\right]^{Nx-N\left(\frac{B}{B-A}\right)} \left[Cx+D(1-x)\right]^{-Nx-N\left(\frac{D}{C-D}\right)}.
\end{eqnarray}
It can also be checked that
\begin{eqnarray}\label{MPSpp}
S''(x)=\frac{BC-AD}{[Ax+B(1-x)][Cx+D(1-x)]}
\end{eqnarray}
is positive over the entire region $0\leq x \leq 1$.
It therefore follows from (\ref{fastmode}), and Eqs.~(\ref{MPrates}), (\ref{MPS}), (\ref{MPSpp}), that
in the bulk, \textit{i.e.} for $N^{-1/2} \ll x \ll 1-N^{-1/2}$, the QSD  reads
\begin{eqnarray}\label{qsdMP}
\pi(x)=\frac{(C-A) (B-D)}{\sqrt{2\pi N(BC\!-\!AD)\,{\cal T}_+(x){\cal T}_-(x)}\,(B\!-\!A\!+\!C\!-\!D)}\;\frac{\left[Ax+B(1\!-\!x)\right]^{Nx-N\left(\frac{B}{B-A}\right)}}{\left[Cx+D(1\!-\!x)\right]^{Nx+N\left(\frac{C}{C-D}\right)}}
\left(\frac{BC-AD}{C\!-\!A\!+\!B\!-\!D}\right)^{\frac{N(BC-AD)}{(B-A)(C-D)}}.
\end{eqnarray}
Moreover, $\pi_1$ and $\pi_{N-1}$ are obtained from Eqs.~(\ref{pi1n}):
\begin{eqnarray}\label{pi1nMP}
\pi_1&=&\sqrt{\frac{N}{2\pi BD(BC-AD)}}\,\frac{(C-A)(B-D)^2}{C-A+B-D}\left(\frac{BC-AD}{C-A+B-D}\right)^{\frac{N(BC-AD)}{(B-A)(C-D)}}B^{-N\left(\frac{B}{B-A}\right)}D^{-N\left(\frac{D}{C-D}\right)}
\nonumber\\
\pi_{N-1}&=&\sqrt{\frac{N}{2\pi AC(BC-AD)}}\,\frac{(C-A)^2(B-D)}{C-A+B-D}\left(\frac{BC-AD}{C-A+B-D}\right)^{\frac{N(BC-AD)}{(B-A)(C-D)}}A^{-N\left(\frac{A}{B-A}\right)}C^{-N\left(\frac{C}{C-D}\right)}.
\end{eqnarray}
%\end{widetext}
Eq.~(\ref{qsdMP}) determines the QSD for ACG evolving according to the fMP [close to the boundaries, one must use Eqs.~(\ref{rec1}) and (\ref{recn}) instead].
Clearly, the resulting QSD is non-Gaussian, which is  especially evident near the tails.
A typical example is shown in Fig.~\ref{qsdmp} where excellent agreement is observed between our analytical results and a numerical solution of the corresponding master equation~(\ref{master}). It is worth noticing that our theory is applicable when $N[S(1)-S(x^*)]\gg 1$ and $N[S(0)-S(x^*)]\gg 1$, which imposes a lower bound on $w$.
Hence, while it is inapplicable in the weak selection limit $Nw \ll 1$, recently investigated by other techniques (see \textit{e.g.}~\cite{Hauert-preprint,Altrock} and references therein), our approach successfully applies to the more general case of finite selection intensity $0<w\leq 1$.

\begin{figure}[ht]
\includegraphics[width=3.2in, height=2.35in,clip=]{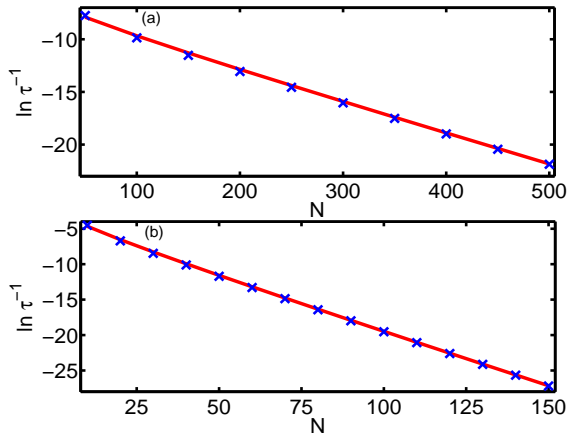}
\caption{{\it (Color online)}. Shown is $\ln{\tau^{-1}}$  as a function of the population size $N$, for $a=0.1$, $b=0.7$, $c=0.6$, $d=0.2$, with $w=0.2$ in panel (a) and $w=0.7$ in panel (b). Excellent agreement is observed  between the analytical solution (solid line), given by Eqs.~(\ref{phimte1}) and (\ref{pi1nMP}), and the numerical solution of the master equation ($\times$'s).} \label{taump}
\end{figure}

\begin{figure}[ht]
\includegraphics[width=3.2in, height=2.35in,clip=]{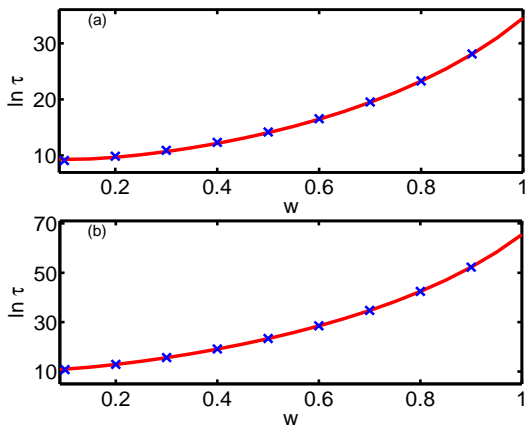}
\caption{{\it (Color online)}. Shown is $\ln{\tau^{-1}}$  as a function of the selection strength $w$, for $a=0.1$, $b=0.7$, $c=0.6$, $d=0.2$, with $N=100$ in panel (a) and $N=200$ in panel (b). Excellent agreement is observed  between the analytical solution (solid line), given by Eqs.~(\ref{phimte1}) and (\ref{pi1nMP}), and the numerical solution of the master equation ($\times$'s).}\label{logMFT_MP}
\end{figure}

The MFTs can now be found by using Eqs.~(\ref{phimte1}) and (\ref{pi1nMP}), with ${\cal T}_-'(0)=|{\cal T}_+'(1)|=1$.
As illustrated in Fig.~\ref{taump}, the unconditional MFT asymptotically exhibits an exponential dependence on the population size $N$. That is, $\tau \propto N^{1/2}e^{N[\Sigma-S(x^*)]}$,  where  $\Sigma\equiv {\rm min}\left[S(0), S(1)\right]$ depends on the entries of the payoff matrix and on the selection intensity $w$.
It follows from Eqs.~(\ref{phimte1}), (\ref{MPS}) and (\ref{pi1nMP}) that
in the biologically relevant case of small (but not too small) selection intensity $N^{-1}\ll w\ll 1$, the MFTs grow exponentially as
$\tau^A\sim N^{1/2}e^{Nw(a-c)^2/[2(c-a+b-d)]}$, $\tau^B\sim N^{1/2}e^{Nw(b-d)^2/[2(c-a+b-d)]},$ and
$\tau=\tau^A \tau^B /(\tau^A + \tau^B)\sim {\rm min}(\tau^A, \tau^B)$. In the opposite limit of large selection strength $w\to 1$, one can show that our results in the leading order coincide with those of Ref.~\cite{Antal}.
For finite selection strength, the exponential dependence of $\tau$ is found to increase monotonically with $w$, as shown in Fig.~\ref{logMFT_MP}.
Note, that in this figure and in all other figures (except Fig.~\ref{exact}), when simulating the master equation~(\ref{master}), the initial number $n$ of $\textsf{A}$'s was chosen to be sufficiently large to avoid immediate fixation prior to reaching the metastable state.

\begin{figure}
\includegraphics[width=3.2in, height=2.35in,clip=]{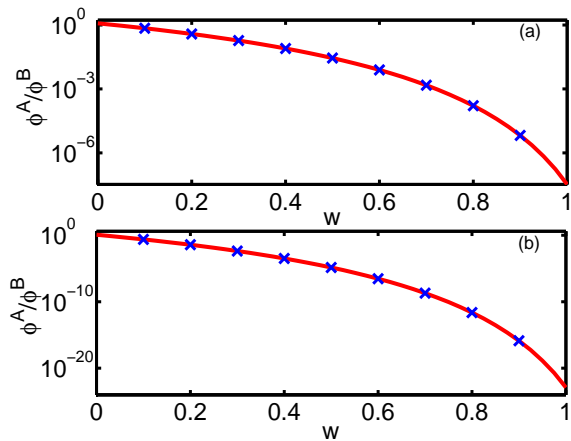}
\caption{{\it (Color online)}. Shown is the ratio $\phi^A/\phi^B$ of the fixation probabilities of species $\textsf{A}$ and $\textsf{B}$
as a function of the selection intensity $w$ in the fMP. The theoretical prediction [Eq.~(\ref{probfixratio})] (solid) is compared with numerical solution ($\times$'s) of the master equation~(\ref{master}). The parameters are $a=0.1, b=0.7, c=0.7$ and $d=0.2$, with $N=100$ in  panel (a) and $N=300$ in (b).} \label{fixratioMP}
\end{figure}

The ratio $\phi^A/\phi^B=\phi^{-1}-1$ allows to assess the influence of selection by comparing
 the fixation probability of species $\textsf{A}$  and $\textsf{B}$ for finite $w$.
It follows from (\ref{mtephi}) that the fixation probability ratio $\phi^A/\phi^B$  is given by
\begin{eqnarray}\label{probfixratio}
\frac{\phi^A}{\phi^B}\!=\!\frac{\pi_{N-1}}{\pi_1}\!=\!\sqrt{\frac{BD}{AC}}\left(\frac{C\!-\!A}{B\!-\!D}\right)
\frac{B^{N\left(\frac{B}{B-A}\right)}D^{N\left(\frac{D}{C-D}\right)}}{A^{N\left(\frac{A}{B-A}\right)}C^{N\left(\frac{C}{C-D}\right)}}\!.
\end{eqnarray}
In Fig.~\ref{fixratioMP}, the asymptotic expression (\ref{probfixratio}) as a function of the selection strength  is compared  with  the numerical solution of the master equation~(\ref{master}), demonstrating an excellent agreement. This figure illustrates the exponential dependence of the ratio $\phi^A/\phi^B$ on $w$ with a marked nonlinear behavior of the exponent. One can see a steep decay as the selection's strength increases (the fixation of $\textsf{B}$ is thus more likely), and for $w$ close to 1 the fixation of $\textsf{A}$ is almost improbable. In the neutral case arising when $w = 0$, the stochastic dynamics is diffusive and the ratio of the fixation probabilities then strongly depends on the initial number $n$ of $\textsf{A}$'s. In stark contrast, for finite $w$ $\,\phi^A/\phi^B$ becomes independent of $n$ (provided that $n\gg 1$~\cite{init}) and converges towards Eq.~(\ref{probfixratio}). This is demonstrated in Fig.~\ref{exact}, where we have compared Eq.~(\ref{probfixratio}) with the numerical solution of the master equation~(\ref{master}) for various initial conditions. Note, that for $w>0$ and $N\to \infty$ the fluxes
to the absorbing states are vanishingly small and (\ref{probfixratio}) becomes singular,
with  $\phi^A/\phi^B \to 0$ or  $\phi^A/\phi^B \to \infty$ depending on the rest of the parameters. When $w\to 0$, one has  $\phi^A/\phi^B \to  x/(1-x)$.

\subsubsection{Linear Moran and local update processes}
The cases of the LMP and LUP, with rates ${\cal T}_{\pm}(x)$ obtained respectively  from Eqs.~(\ref{LMP}) and (\ref{LU}), can be studied in the same manner as the fMP.
Given the birth and death rates ${\cal T}_{\pm}(x)$, one obtains the action [see Eq.~(\ref{Ssingle})] and, with Eqs.~({\ref{fastmode}), (\ref{pi1n}), and (\ref{phimte1}), one can calculate the QSD, fixation probability and MFTs.
This  leads essentially to  the same qualitative features as in the fMP with low selection intensity $w$.
Our findings are summarized in Fig.~\ref{tauLU}, where our prediction for the unconditional MFT for the LUP is found to grow exponentially with $N$ and $w$, in excellent agreement with numerical results

\subsubsection{Fermi process}
We  now consider the FP
that has recently received considerable attention  (see e.g.~\cite{Traulsen2006,Altrock,Hauert-preprint}).
As above,  the transition rates ${\cal T}_{\pm}(x)$ for the FP are obtained
from Eq.~(\ref{FP}). With Eq.~(\ref{Ssingle}), the action in this case is quadratic
\begin{eqnarray}\label{FPPact}
S(x)=\int^x w[(c-a)q+(d-b)(1-q)]dq=wx(d-b)\left[1-x/(2x^*)\right]\,,
\end{eqnarray}
and $S''(x)=w(c-a+b-d)>0$. Using Eqs.~(\ref{pi1n}) and (\ref{FPPact}),
after some algebra one obtains the following expressions for $\pi_1$ and $\pi_{N-1}$:
\begin{eqnarray}\label{FPpi}
\pi_1&=&\sqrt{\frac{Nw}{2\pi}}\frac{(c-a)(b-d)}{(c-a+b-d)^{3/2}}\sinh[(b-d)w]e^{\frac{(b-d)w}{2}}\exp\left[-\frac{wN(b-d)^2}{2(c-a+b-d)}\right]\,,\nonumber\\
\pi_{N-1}&=&\sqrt{\frac{Nw}{2\pi}}\frac{(c-a)(b-d)}{(c-a+b-d)^{3/2}}\sinh[(c-a)w]e^{\frac{(c-a)w}{2}}\exp\left[-\frac{wN(c-a)^2}{2(c-a+b-d)}\right]\,.
\end{eqnarray}
It follows from these results that in the case of the FP, our approach is valid as long as $w \gg N^{-1}$,
with $N\gg 1$. Therefore, our results are complementary to those of earlier works on this model which were carried out in the weak selection limit by treating selection as a linear perturbation to the neutral case $w=0$ (see, \textit{e.g.}, Refs.~\cite{Traulsen2006,Altrock,Hauert-preprint}).

\begin{figure}[ht]
\includegraphics[width=3.2in, height=2.35in,clip=]{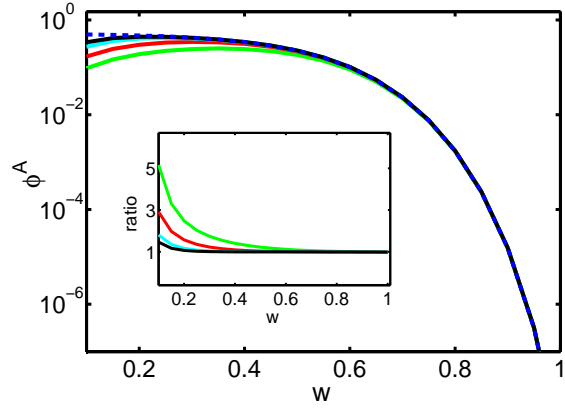}
\caption{{\it (Color online)}. Shown is the fixation probability of the $\textsf{A}$ species versus the selection intensity $w$ for the fMP. The dashed line is the analytical prediction given by Eqs.~(\ref{mtephi}) and (\ref{pi1n}). The four solid lines are numerical solutions of the master equation~(\ref{master}) starting with $n=5,10,20,30$ (bottom to top) initial $\textsf{A}$'s.
 The numerical results are found to converge towards the analytical prediction when $w$ increases, with a convergence
that improves when $n$ increases. Inset: Ratios of the above four numerical curves
and the analytical prediction (top to bottom). One clearly observes that the smaller $w$ is, the larger $n$ must be
to avoid fixation prior to reaching the coexistence state. The parameters are $a=0.1$, $b=0.7$, $c=0.6$, $d=0.2$ and $N=300$.} \label{exact}
\end{figure}

\begin{figure}[ht]
\includegraphics[width=3.2in, height=2.35in,clip=]{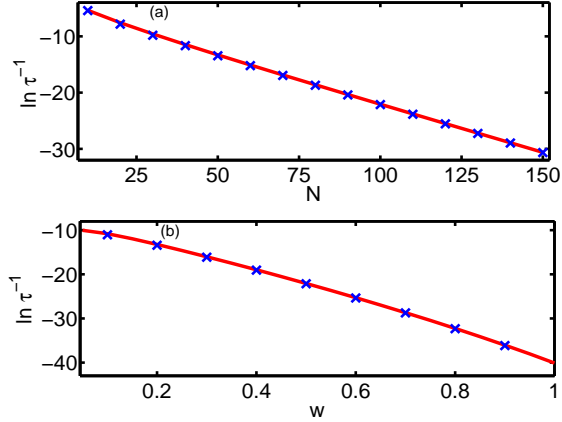}
\caption{{\it (Color online)}. (a) For the LUP (\ref{LU}), shown is $\ln \tau^{-1}$ as a function of
the population size $N$  and parameters $w=0.5$, $a=0.2$, $b=c=0.9$, and $d=0.1$.
In (b), shown is $\ln \tau^{-1}$ for the LUP as a function of the selection intensity $w$, for the same process and
same parameters as in (a) but $N=100$. An excellent agreement between our analytical predictions (solid lines) and the numerical solution of the master equation ($\times$'s) is observed.} \label{tauLU}
\end{figure}

Using Eqs.~(\ref{fastmode}), (\ref{phimte1}), (\ref{FPPact}) and (\ref{FPpi}) one can explicitly obtain the QSD, the fixation probability and the (unconditional and conditional) MFTs. Here, one obtains the following asymptotic behavior of the conditional MFTs:
\begin{eqnarray}
\tau^A\sim N^{1/2} \exp\left[\frac{wN(c-a)^2}{2(c-a+b-d)}\right]\;,\;\;\;\;\tau^B\sim N^{1/2} \exp\left[\frac{wN(b-d)^2}{2(c-a+b-d)}\right],
\end{eqnarray}
while the unconditional MFT satisfies $\tau=\tau^A \tau^B/(\tau^A +\tau^B)$. We thus notice that, to leading order, the MFTs for the FP coincide with those obtained for the fMP in the limit of small (but not too small) selection strength $N^{-1}\ll w \ll 1$.
In addition to the MFTs, an interesting quantity to compute is the ratio of the fixation probabilities $\phi^A$ and $\phi^B$
\begin{equation}\label{fixratio}
\frac{\phi^A}{\phi^B}=\frac{\sinh[(c-a)w]}{\sinh[(b-d)w]}\exp\left[\frac{(N-1)w(a+b-c-d)}{2}\right]\,.
\end{equation}
This ratio is larger than unity  if $b-d>c-a$, i.e. when $x^*>1/2$, which simply means that the closer $x^*$ is to $1$, the easier it is for species $\textsf{A}$ to fixate. In Fig.~\ref{probfix}, we show the ratio between $\phi^A$ and $\phi^B$ given by Eq.~(\ref{fixratio}) as a function of $N$ for a
low and high selection strength ($w=0.2$ and $w=0.7$, respectively) and find excellent agreement with the  numerical solution of the master equation~(\ref{master}).
One can easily show that our asymptotic result~(\ref{fixratio}) coincides in the leading order with the exact result found from Eq.~(\ref{ProbFix}) (which takes a simple form in this case), where the difference in the pre-factor stems from self-interaction terms that were not excluded in our treatment [see Eq.~(\ref{FP})]~\cite{o1}.

\begin{figure}[ht]
\includegraphics[width=3.2in, height=2.35in,clip=]{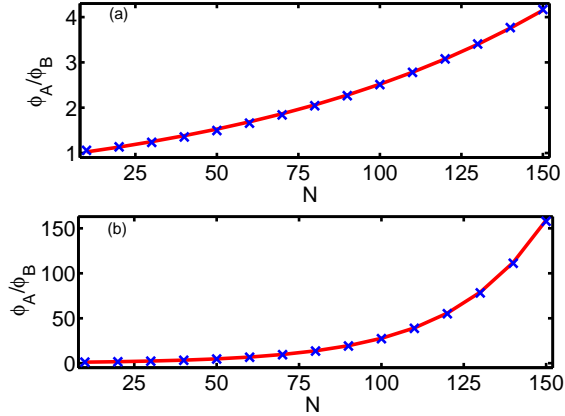}
\caption{{\it (Color online)}. The ratio $\phi^A/\phi^B$ as a function of $N$ for the FP~(\ref{FP}). Comparison between the analytical result given by Eq.~(\ref{fixratio}) (solid line) and the numerical solution  of the master equation~(\ref{master}) ($\times$'s).
The parameters in (a) are $a=0.5$, $b=c=2$, $d=0.4$, \textit{i.e.} $x^*\simeq 0.516$, and $w=0.2$, and in (b)
$a=0.3$, $b=c=1$, $d=0.2$, \textit{i.e.} $x^*\simeq 0.533$, and  $w=0.7$. As  $x^*>1/2$, we notice that the
fixation probability of species  $\textsf{A}$ increases with $w$ and $N$.} \label{probfix}
\end{figure}

\section{Coordination games: fixation probability}
In this section we use the WKB method to asymptotically compute the fixation probability in CG for $N\gg 1$ and finite $w$. After the presentation of the general treatment, our theoretical results  are applied to the fMP~(\ref{MP}) and FP~(\ref{FP}) update rules, and are compared with those obtained from the FPE.

\subsection{General treatment \& results}
In CG, the elements of the payoff matrix satisfy $d>b$ and $a>c$. Thus, within the realm of rate equations, the fixed point $x^*$ (\ref{IntFP}) is a repellor, whereas the absorbing states $x=0$ and $x=1$ are attractors. In the presence of noise, the fixation of the intruding species occurs rapidly~\cite{Antal} and therefore the
main interest is in the fixation probability $\phi^A_n$: the probability that starting from $n<n_*$ individuals of type $\textsf{A}$, the species $\textsf{A}$ will fixate the population. In terms of the transition rates $T^{\pm}(n)$, $\phi_n^{A}$ obeys the following difference equation~\cite{Gardiner,vanK,Nowak,Antal}
\begin{equation}\label{diffeq}
T^{+}(n)\phi_{n+1}^{A}+T^{-}(n)\phi_{n-1}^{A}-[T^{+}(n)+T^{-}(n)]\phi_n^{A}=0\,,
\end{equation}
with  boundary conditions $\phi_0^{A}=0,\phi_N^{A}=1$. Here, the probability $\phi_n^A$ that the $\textsf{A}$'s fixate starting from $n$ individuals of type
$\textsf{A}$ is given by a sum of two components. The first is the probability to fixate starting from $n+1\,$ $\textsf{A}$'s multiplied by the probability to jump to state $n+1$ from state $n$, which is $T^{+}(n)/[T^{+}(n)+T^{-}(n)]$. The second component is the probability to fixate starting from $n-1\,$ $\textsf{A}$'s multiplied by the probability to jump to state $n-1$ from state $n$, which is $T^{-}(n)/[T^{+}(n)+T^{-}(n)]$. Note, that in this section (and differently from the treatment of ACG), as there is no metastability, the results strongly depend on the initial condition, that is, on the initial number of $\textsf{A}$'s.

As $\phi^A_n\equiv \phi^A(x)$ is a cumulative distribution function, it is more convenient to consider the corresponding PDF, defined as
${\cal P}_n\equiv {\cal P}(x)\equiv \phi^A_{n+1}-\phi^A_n$. ${\cal P}(x)$ is peaked in the vicinity of $x^*=n_*/N$ (see insets of Figs.~\ref{coordw})
and can be shown to satisfy ${\cal P}_0=\phi^A_1$, ${\cal P}_{N-1}=1-\phi^A_{N-1}$ and $\sum_0^{N-1}{\cal P}_n=1$.
Rewriting Eq.~(\ref{diffeq}) as ${\cal T}_{+}(x)[\phi^A(x+N^{-1})-\phi^A(x)]-{\cal T}_{-}(x)[\phi^A(x)-\phi^A(x-N^{-1})]=0$ and using the definition of ${\cal P}(x)$, one obtains the following equation for ${\cal P}(x)$
\begin{equation}\label{diffeq2}
{\cal T}_{+}(x){\cal P}(x)-{\cal T}_{-}(x){\cal P}(x-N^{-1})=0\,.
\end{equation}
This equation is similar in form to the QSME~(\ref{masterqsd}) and is treated using the WKB ansatz
\begin{equation}\label{picoord}
{\cal P}(x)={\cal A}_{{\rm CG}} \, e^{-N{\cal S}(x)-{\cal S}_1(x)}.
\end{equation}
Substituting (\ref{picoord}) into (\ref{diffeq2}), one obtains in the leading ${\cal O}(1)$ order
${\cal T}_{+}(x)-{\cal T}_{-}(x)e^{{\cal S}'(x)}=0$, whose solution reads
\begin{equation}\label{scoord}
{\cal S}(x)=-S(x)=\int^x \ln [{\cal T}_+(\xi)/{\cal T}_-(\xi)]d\xi.
\end{equation}
Here,  ${\cal S}(x)$ is the negative of the expression (\ref{Ssingle}), and thus $S''(x^*)<0$. In the subleading ${\cal O}(1/N)$ order, one obtains ${\cal S}_1(x)=(1/2){\cal S}'(x)=(1/2)\ln [{\cal T}_+(x)/{\cal T}_-(x)]$. As in Sec.~II, the constant ${\cal A}_{{\rm CG}}$ in Eq.~(\ref{picoord}) is found by normalizing the Gaussian asymptote of ${\cal P}(x)$  in the vicinity of $x^*$. With Eq.~(\ref{scoord}), the final result for ${\cal P}(x)$ reads
\begin{equation}\label{pdfcoord}
{\cal P}(x)=\sqrt{\frac{| S''(x^*)|}{2\pi N}}\sqrt{\frac{{\cal T}_-(x)}{{\cal T}_+(x)}} e^{N[S(x)-S(x^*)]}.
\end{equation}
As shown in the appendix, this result coincides up to subleading-order corrections with the exact solution of Eq.~(\ref{diffeq2}). In particular, this implies that the recursion solution of Eq.~(\ref{diffeq2}) near the boundaries $x=0$ and $x=1$ exactly coincides with the
WKB result at $x\ll 1$ and $1-x\ll 1$, respectively, and  no special treatment is thus required near those boundaries (see also Ref.~\cite{AM}).
With Eq.~(\ref{pdfcoord}), we can write down the fixation probability, $\phi_n^A=\sum_{i=0}^{n-1}{\cal P}_i$, as
\begin{equation}\label{wkbsum}
\phi_n^A=\sqrt{\frac{| S''(x^*)|}{2\pi N}}\sum_{i=0}^{n-1}\sqrt{\frac{T^-(i)}{T^+(i)}}\;e^{N[S(i/N)-S(x^*)]}.
\end{equation}
This expression gives the fixation probability of species $\textsf{A}$ for any finite $w$ (see Figs.~\ref{coordw}), where its accuracy holds in the entire region of $x$ including the boundaries.

Remarkably, the summation in Eq.~(\ref{wkbsum}) can be drastically simplified for $x\ll x^*$, \textit{i.e.}, $n\ll n_*$. This corresponds to the biologically important limit of the fixation probability of a few intruders of type  $\textsf{A}$ in a sea of $\textsf{B}$'s~\cite{NowakTaylor,Nowak,Hauert-preprint}.
It is now convenient to split our discussion into two cases. For small (but not too small) selection intensity $N^{-1}\ll w \ll 1$, ${\cal P}(x)$ is slowly varying, and the sum~(\ref{wkbsum}) can be safely transformed into an integral. In this case, the term $\sqrt{{\cal T}_-(x)/{\cal T}_+(x)}\simeq 1$ can
be omitted from the integration, yielding
\begin{eqnarray}\label{fixsmallw}
\hspace{-2.5mm}\phi^A(x)\simeq  \sqrt{\frac{N| S''(x^*)|}{2\pi}}\int_0^{x}e^{N[S(\xi)-S(x^*)]}d\xi.
\end{eqnarray}
This approximation is valid when $|{\cal P}_{n+1}/{\cal P}_n-1|\ll 1$, \textit{i.e.}, when $|S'(x)|\ll 1$,
which assures that ${\cal T}_+(x)\simeq {\cal T}_-(x)$ and holds excellently for $w\ll 1$.

In the second case, $w={\cal O}(1)$, the sum~(\ref{wkbsum}) is dominated by its last term when $1\ll n\ll n_*$.
In this case, denoting  $k=n-1-i$, one can Taylor-expand the summand about $i=n$ in Eq.~(\ref{wkbsum}), yielding
$e^{-N{\cal S}(i/N)-{\cal S}_1(i/N)}\simeq e^{-N{\cal S}(n/N)-{\cal S}_1(n/N)+(k+1){\cal S}'(n/N)+{\cal O}(1/N)}.$
Plugging this expression into the sum~(\ref{wkbsum}), one has [with ${\cal S}'(x)=-S'(x)$]
\begin{eqnarray}\label{fixlargew}
\phi^A(x)\simeq {\cal P}(x)\sum_{k=0}^{n-1}e^{-(k+1)S'(x)}\simeq \frac{{\cal P}(x)}{e^{S'(x)}-1},
\end{eqnarray}
where $S'(x)>0$ for $x<x^*$, and we have replaced the upper limit of the sum by infinity. Results~(\ref{fixsmallw}) and (\ref{fixlargew}) are valid for $N^{-1}\ll x \ll x^*$. Clearly, similar approximations can be made near $x=1$, in the region  $N^{-1}\ll 1-x\ll 1$.

In Fig.~\ref{coordw}, for the fMP, we compare the numerical results for $\phi^A$ with the WKB solution
(\ref{wkbsum}) and with its approximation (\ref{fixlargew}), and an excellent agreement is observed.
 The latter improves as $w$ is increased. From (\ref{fixlargew}), we infer that the fixation probability is exponentially small
for finite selection intensity and therefore one generally has  $\phi^A(x)<x$ when   $x\ll 1$. In stark contrast with the weak selection limit~\cite{NowakTaylor,Nowak,Pacheco,Altrock,Hauert-preprint},  this implies that when $w$ is finite,
selection {\it always} opposes the replacement of the wild species ($\textsf{B}$) by  mutants ($\textsf{A}$).

\begin{figure}[ht]
\includegraphics[width=3.2in, height=2.35in,clip=]{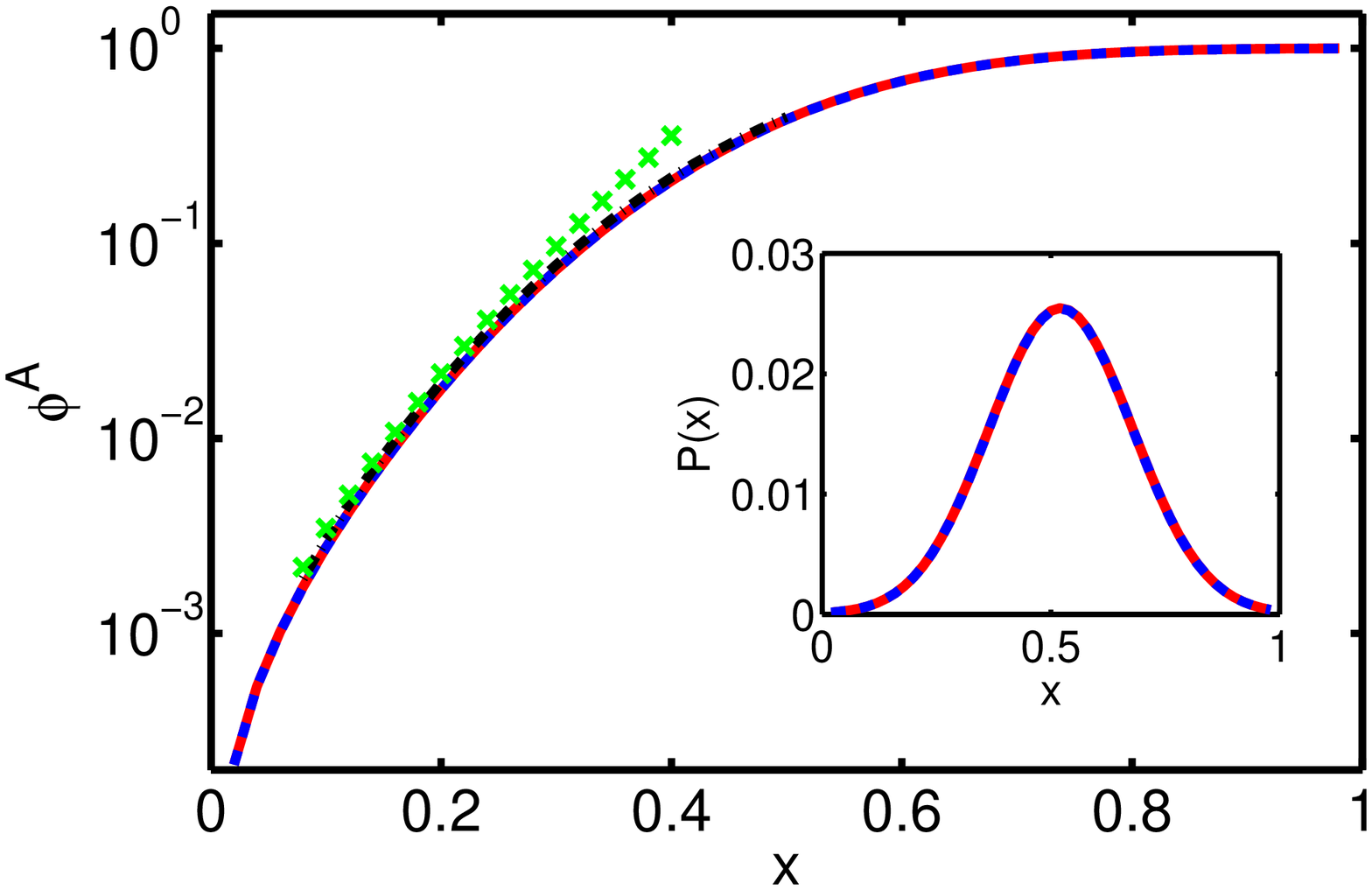}
\includegraphics[width=3.2in, height=2.35in,clip=]{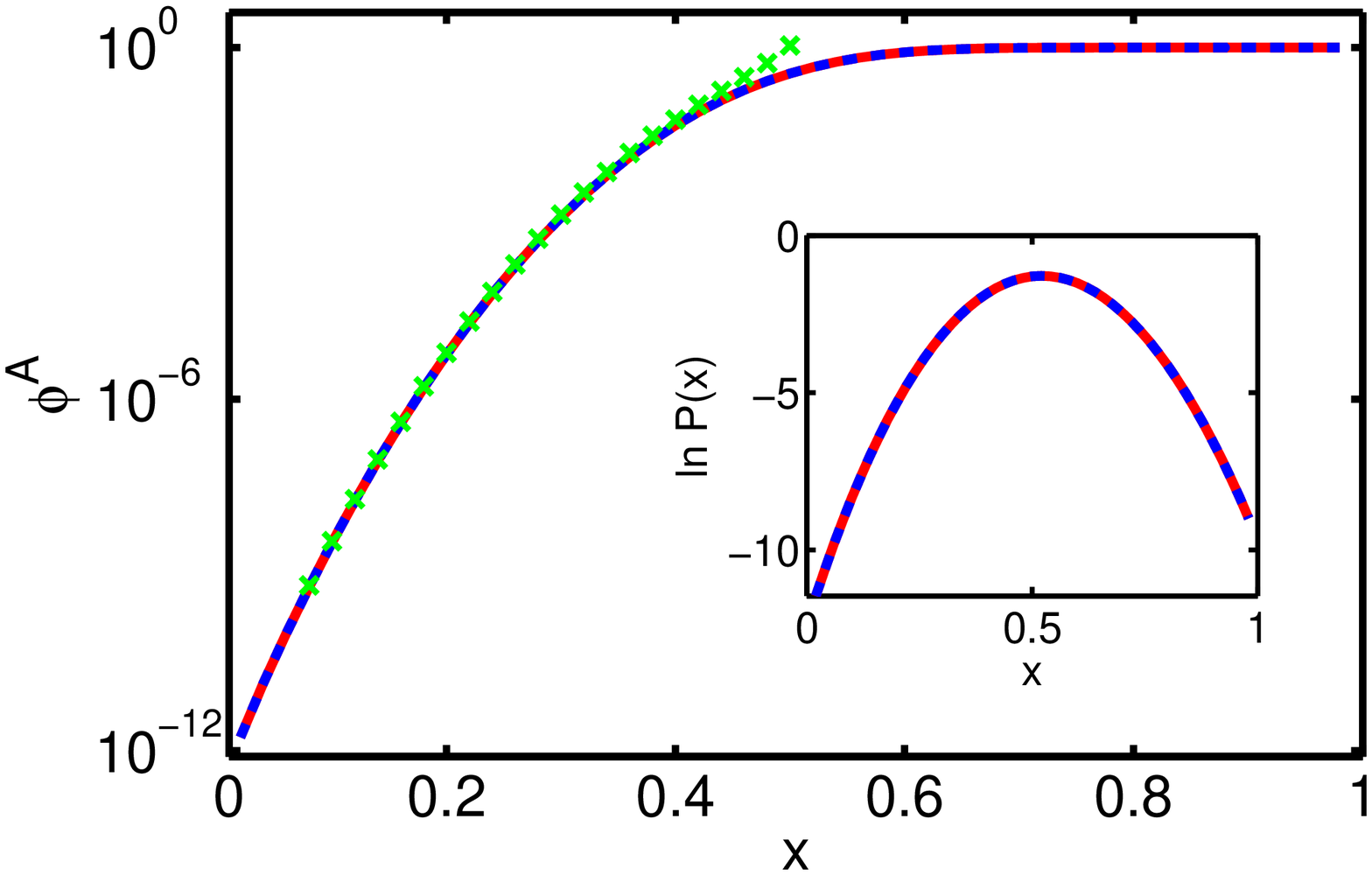}
\caption{{\it (Color online)}. Fixation probability of species $\textsf{A}$ evolving according to the fMP~(\ref{MP}): comparison between the expression (\ref{wkbsum}) (dashed), its approximation (\ref{fixlargew}) ($\times$'s), and numerical results (solid) as  function of $x$. Note that the theoretical (\ref{wkbsum}) and numerical results are indistinguishable.
The parameters are $N=100$, $a=1.2$, $b=0.1$, $c=0.3$, $d=1.1$, with $w=0.2$ and $w=0.7$ in the left and right panels,
 respectively. The quality and range of validity of the approximation~(\ref{fixlargew}) improves as $w$ is increased. In the left panel
the small-$w$ approximation~(\ref{fixsmallw}) is also reported (dash-dotted). In this case, as $S'(x)\ll 1$, the approximation~(\ref{fixsmallw}) is superior to (\ref{fixlargew}), see text. In the insets of both panels, we show a comparison between
the analytical [Eq.~(\ref{pdfcoord})] (dashed) and numerical (solid) results for ${\cal P}(x)$, and excellent agreement is found over the entire range $0\leq x \leq 1$. }
\label{coordw}
\end{figure}

\subsection{Applications}
We now apply the above general results to the cases of  CG evolving according to the fMP~(\ref{MP}) and FP~(\ref{FP}) and compare our theoretical results with those of the FPE. For the fMP, the rates are given by (\ref{MPrates}), with $A>C$ and $D>B$. In this case the action $S(x)$ is
given by Eq.~(\ref{MPS}). The fixation probability of species $\textsf{A}$ starting with $n=Nx \ll n_*$ individuals, is given by
Eqs.~(\ref{fixsmallw}) and (\ref{fixlargew}), i.e.
\begin{eqnarray}\label{fixAMP}
\phi^A(x)&\simeq& N\int_0^{x}{\cal P}_{{\rm fMP}}(\xi)d\xi \;\;\;\;\;\mbox{for}\;\; N^{-1}\ll w\ll 1\nonumber\\
\phi^A(x)&\simeq& \frac{{\cal P}_{{\rm fMP}}(x)}{e^{S'(x)}-1}\hspace{1.5cm}\mbox{for}\;\;w={\cal O}(1), \label{fixMP}
\end{eqnarray}
where ${\cal P}_{{\rm fMP}}(x)$ is given by Eq.~(\ref{pdfcoord}) and $S(x)$
given by (\ref{MPS}). In Fig.~\ref{coordw}, we compare (for $w=0.2$ and $w=0.7$) the theoretical predictions
 [Eq.~(\ref{fixMP})] with the numerical results and find an excellent agreement over the entire range $0<x<1$.
Results~(\ref{fixAMP}) generalize  the results of Ref.~\cite{Antal} by considering arbitrary (finite) selection strength $0<w\leq 1$
and by including the subleading-order correction.

\begin{figure}[ht]
\includegraphics[width=3.2in, height=2.35in,clip=]{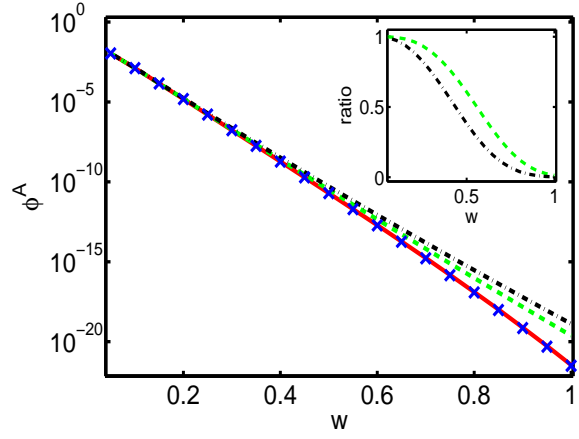}
\caption{{\it (Color online)}.
Fixation probability  $\phi^A(x)$  evolving according to the fMP~(\ref{MP}) versus $w$: comparison between the WKB result
[Eq.~(\ref{fixAMP})] (solid), results of the full [Eqs.~(\ref{fixmp}) and (\ref{fixfp})] (dashed) and linearized [Eqs.~(\ref{fixmp}) and (\ref{fixfp_l})] (dash-dotted) FPE, and numerical solution of Eq.~(\ref{master}) ($\times$'s). Parameters are $a=2$, $b=0.2$, $c=0.3$, $d=1.8$, $N=150$, with initial condition $x=n/N=10/150$. The WKB solution agrees excellently with the numerical results, while the FPE approximations
systematically deviate from the WKB result as $w$ increases. Inset: ratios between the WKB and the results of the
full (dashed), and linearized (dash-dotted) FPE as  function of $w$. The linearized FPE has a narrower region of applicability, see text.}\label{fokkerw}
\end{figure}

For the FP (\ref{FP}) starting with $n\ll n_*$ mutants, Eqs.~(\ref{fixsmallw}) and (\ref{fixlargew}) can be
explicitly calculated. Using (\ref{FPPact}) one finds
\begin{eqnarray}\label{fixAFP}
\phi^A(x)&=&\frac{1}{2}\left\{\mbox{erf}\left[\sqrt{Nx^*\alpha}(x/x^*-1)\right]+\mbox{erf}(\sqrt{Nx^*\alpha})\right\}\;\;\mbox{for}\;\;\; N^{-1}\ll w\ll 1\nonumber\\
\phi^A(x)&=&\sqrt{\frac{\alpha}{\pi Nx^*}}\;\frac{e^{-\alpha(x/x^*-1)\left[N(x-x^*)+1\right]}}{e^{-2\alpha(x/x^*-1)}-1}\hspace{2.0cm}\mbox{for}\;\;w={\cal O}(1).
\end{eqnarray}
where ${\rm erf}(x)=(2/\sqrt{\pi})\int_{0}^{x} e^{-y^2} dy$ denotes the usual error function, and $\alpha=w(d-b)/2>0$.
The small-$w$ result coincides with the exact results in the continuum limit (see, e.g., Refs.~\cite{Traulsen2006,Hauert-preprint}).

Fixation probabilities are often computed using diffusion approximations, like the FPE~\cite{Kimura,Ewens,Pacheco,Hauert-preprint},
that can be obtained from a van Kampen size expansion. This expansion implicitly assumes that  $\phi^A(x)$ varies slowly over the entire range of $0<x<1$~\cite{vanK,Gardiner}. Here, we are interested in comparing the predictions of the FPE with those of the WKB approach
when the selection  intensity is small (but not too small) $N^{-1}\ll w \ll 1$. In this case, the fixation probability in both the WKB and Fokker-Planck treatments~\cite{Ewens,Traulsen,Hauert-preprint}, is conveniently rewritten as
\begin{eqnarray}\label{fixmp}
\phi^A(x)=\frac{\Psi(x)}{\Psi(1)},\;\mbox{where}\;\Psi(x)=\int_0^x e^{-\int_0^{\xi} \Theta(z)dz}d\xi.
\end{eqnarray}
Here, by comparing Eq.~(\ref{fixmp}) to Eq.~(\ref{fixsmallw}), one has $\Theta_{\rm WKB}(x)=N\ln{\left[{\cal T}_+(x)/{\cal T}_-(x)\right]}$, while
\begin{eqnarray}\label{fixfp}
\Theta_{\rm FPE}(x)=2N\; \left[\frac{{\cal T}_+(x) - {\cal T}_-(x)}{{\cal T}_+(x) + {\cal T}_-(x)}\right].
\end{eqnarray}
Furthermore, the FPE is often considered within the linear noise approximation~\cite{Gardiner,vanK,Pacheco}. In this case, $\Theta_{\rm FPE}(x)$ (\ref{fixfp}) is expanded to linear order in $x-x^*$, yielding
\begin{eqnarray}\label{fixfp_l}
\Theta_{\rm  \ell FPE}(x)=2N(x-x^*)\; \left[\frac{{\cal T}'_+(x^*) - {\cal T}'_-(x^*)}{{\cal T}_+(x^*) + {\cal T}_-(x^*)}\right].
\end{eqnarray}
To see how the WKB approximation compares with that of the FPE, we Taylor expand the functions $\Theta$  around $x^*$,
and compute $\Delta\Theta_{\rm   FPE}(x)=\Theta_{\rm WKB}(x)-\Theta_{\rm   FPE}(x)$ and $\Delta\Theta_{\ell{\rm   FPE}}(x)=\Theta_{\rm WKB}(x)-\Theta_{\ell{\rm   FPE}}(x)$. For the fMP one finds
\begin{eqnarray}\label{diff1}
\Delta\Theta_{\rm   FPE}(x)\simeq C_{\rm   FPE}\, N w^3(x-x^*)^3 \;,\;\;\;\;\Delta\Theta_{\rm \ell   FPE}(x)\simeq C_{\rm \ell   FPE}\, N w^2(x-x^*)^2 ,
\end{eqnarray}
where $C_{\rm   FPE} = (1/12)(a-b-c+d)^6/[(a-b-c+d)(1-w)+(ad-bc)w]^3$, and $C_{\rm \ell   FPE}=(1/2)(a-b+c-d)
(a-b-c+d)^3/[(a-b-c+d)(1-w)+(ad-bc)w]^2$, are both ${\cal O}(1)$. Results~(\ref{diff1}) demonstrate
that the exponent $\Theta$ of our theory and those obtained from the FPE significantly deviate from each other
when $x-x^*={\cal O}(1)$ [e.g. when $x\ll 1$  or $1-x\ll 1$] and  $w$ is finite. Looking at Eq.~(\ref{diff1}), the results of the full and linear FPE agree with the WKB theory and numerical calculations (see also Figs.~\ref{fokkerw} and \ref{fokkerN}) when $N^{-1}\ll w \ll N^{-1/3}$ and $N^{-1}\ll w \ll N^{-1/2}$, respectively.
This implies that demanding that $w\ll 1$ does not guarantee the applicability of the FPE, as the results of the full and linear FPE are plagued by
exponentially large errors already when $w \gtrsim N^{-1/3}$ and $w \gtrsim N^{-1/2}$, respectively.
While the predictions of the FPE further deteriorate when  $w$ increases, our theory improves and allows the accurate calculation of the exponentially
small fixation probability for any finite value of $w$ (see Fig.~\ref{fokkerw})\cite{fermi}.

\begin{figure}[ht]
\includegraphics[width=3.2in, height=2.35in,clip=]{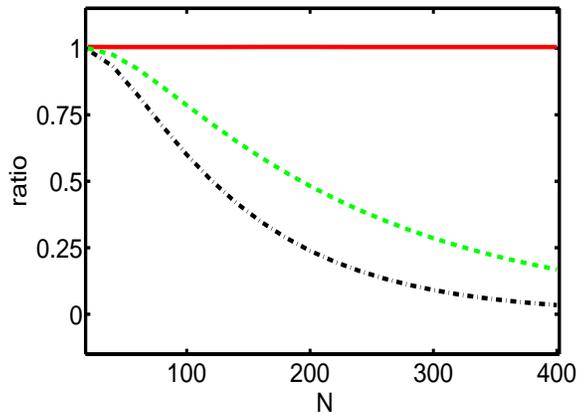}
\caption{{\it (Color online)}.
Shown are the ratios of the WKB result for $\phi^A(x)$ [Eq.~(\ref{fixAMP})] with the numerical result (solid), with the full FPE [Eqs.~(\ref{fixmp}) and (\ref{fixfp})] (dashed), and with the linearized FPE [Eqs.~(\ref{fixmp}) and (\ref{fixfp_l})] (dash-dotted), as a function of $N$, for the fMP~(\ref{MP}). Here $x=10/N$ (i.e. $n=10$),  $a=2$, $b=0.2$, $c=0.3$, $d=1.8$ and $w=0.5$. The WKB predictions agree excellently with the numerical results, while
there are systematic deviations (that increase with $N$) between the latter and the predictions of the
the diffusion approximations (full and linearized FPE), see text.}\label{fokkerN}
\end{figure}

\section{Summary and Conclusion}
In this work, we have studied large-fluctuation-induced fixation in  ($2 \times 2$) anti-coordination and coordination  evolutionary games using a WKB-based approach.
In both classes of games, the deterministic description is characterized by the existence of an interior fixed point separating two absorbing fixed points. The latter are the only possible outcomes of the dynamics when internal stochasticity is taken into account. Yet, for ACG, the
mean time necessary to reach one of the absorbing fixed points (mean fixation time) is typically very large as the
system typically spends an exponentially long time in the metastable coexistence state. On the other hand, in CG fixation occurs rapidly, and the quantity of interest is the (small) probability that a system comprising of a few mutants takes over the entire population, causing the extinction of the wild species.
As the stochastic dynamics of evolutionary games is formulated in terms of one-dimensional single-step birth-death processes
(with frequency-dependent rates), there exist exact formulas for the mean time and probability of fixation. However, these unwieldy
expressions are nontrivial to analyze and cannot be generalized to multi-step processes or higher dimensions. To circumvent this difficulty, one popular approach is to resort to diffusion approximations, as those based on the FPE. However, the latter are ill-suited to describe phenomena like fixation that  result from large fluctuations.

Here we have presented an alternative approach based on the WKB theory, which we have generalized to account for the existence of multiple absorbing states. Within this approach, the stochastic dynamics of the system is formally mapped, in the leading order, onto a Hamiltonian system, whose nontrivial zero-energy trajectory encodes, with the maximum probability, the rare event in question. By using the WKB approach complemented by recursive solutions of the master equation near the two absorbing boundaries, we have obtained general results, for the complete statistics including large fluctuations, of generic one-dimensional birth-death systems possessing two absorbing states. Our results were obtained including important pre-exponential factors, which were found to scale as some power of the typical population size. Along with the generic treatment, we have also considered the most frequently used
microscopic dynamics, based on the frequency-dependent Moran process (fMP), the linear Moran and local update processes, as well as the Fermi process (FP). In particular we have focused on the fMP and FP and have obtained explicit analytical results for the complete metastable probability distribution function of population sizes and for the fixation probability and times in ACG, as well as for the fixation probability in CG. All these results were obtained for arbitrary and finite $w$, which allowed us to shed further light on the combined influence of selection and random fluctuations in evolutionary processes.
Finally, by comparing our analytical results to those of the FPE, we have explicitly found the region of applicability of the FPE and have shown that the WKB approach is vastly superior over the FPE when the selection strength is finite.

While  we have here focused on $2\times 2$ evolutionary games, the WKB-based method presented in this work
can be generalized and is expected to be useful for multi-species problems, like the rock-paper-scissors games
that have recently received considerable attention (see, \textit{e.g.}, ~\cite{Hofbauer,RPS}).

\section*{Appendix}
\renewcommand{\theequation}{A\arabic{equation}}
\setcounter{equation}{0}

In this appendix, we show that the WKB result (\ref{pdfcoord}) asymptotically coincides with the exact solution of (\ref{diffeq2}). Indeed by using recursion, the exact solution of (\ref{diffeq2}) reads
\begin{eqnarray}\label{wp}
\hspace{-3mm}{\cal P}_n^{{\rm exact}}={\cal P}_0\prod_{i=1}^n \left(\frac{T^-(i)}{T^+(i)}\right)={\cal P}_0
\exp\left[\sum_{i=1}^n \ln \left(\frac{T^-(i)}{T^+(i)}\right)\right].
\end{eqnarray}
Here, ${\cal P}_0$ is a constant to be found by normalization. For $N\gg 1$, the sum in the exponent of Eq.~(\ref{wp}) can be transformed into an integral using Euler-Maclaurin formula $\,\sum_{i=0}^n f(i)=\int_0^n f(x)dx+(1/2)[f(n)+f(0)]+(1/12)[f'(n)+f'(0)]-\dots.$ As the sum is in the exponent of (\ref{wp}),
such a transformation should be done carefully and subleading-order corrections to the integral must be taken into account. Therefore,
 for $N\gg 1$ and $x=n/N$, one has
\begin{eqnarray}\label{sum}
\sum_{i=1}^{n}\ln \left(\frac{T^-(i)}{T^+(i)}\right)\simeq N\int_0^x \ln \left(\frac{T_-(\xi)}{T_+(\xi)}\right) d\xi+\frac{1}{2}\ln \left(\frac{{\cal T}_-(x)/{\cal T}_+(x)}{{\cal T}_-^{\prime}(0)/{\cal T}_+^{\prime}(0)}\right),
\end{eqnarray}
up to ${\cal O}(1/N)$ corrections, where we used the fact that ${\cal T}_-(0)/{\cal T}_+(0)={\cal T}_-^{\prime}(0)/{\cal T}_+^{\prime}(0)$. Thus, Eq.~(\ref{wp}) becomes
\begin{equation}\label{wp1}
{\cal P}^{{\rm exact}}(x)\simeq {\cal P}_0 \sqrt{\frac{{\cal T}_-(x)/{\cal T}_+(x)}{{\cal T}_-^{\prime}(0)/{\cal T}_+^{\prime}(0)}} e^{N[S(x)-S(0)]}.
\end{equation}
where we have used the definition of $S(x)$ from Eq.~(\ref{scoord}). Finally, ${\cal P}_0$ is determined by demanding that $\sum_0^{N-1}{\cal P}_n=1$.
As before, for $N\gg 1$ we approximate this sum by an integral, whose main contribution arises from the
Gaussian region around $x\simeq x^*$, where the function ${\cal P}(x)$ varies slowly. By doing so, one obtains
\begin{equation}\label{wp0}
{\cal P}_0=\sqrt{\frac{|S''(x^*)|}{2\pi N}}\sqrt{\frac{{\cal T}_-^{\prime}(0)}{{\cal T}_+^{\prime}(0)}}e^{N[S(0)-S(x^*)]}.
\end{equation}
With this result, (\ref{wp1}) coincides with Eq.~(\ref{pdfcoord}) obtained directly from our WKB treatment.


\begin{thebibliography}{99}
\bibitem{Hofbauer} J. Hofbauer and K. Sigmund, {\it Evolutionary Games and Population Dynamics} (Cambridge University Press, Cambridge, 1998).
\bibitem{Szabo} G. Szab\'o and G. F\'ath, Phys. Rep. {\bf 446}, 97 (2007)
\bibitem{Nowak} M. A. Nowak, {\it Evolutionary Dynamics} (Belknap Press, 2006).
\bibitem{Gintis} H. Gintis, {\it Game Theory Evolving} (Princeton University Press, Princeton, 2000).
\bibitem{Maynard} J. Maynard Smith, {\it Evolution and the Theory of Games} (Cambridge University Press, Cambridge, England, 1982).
\bibitem{Hamilton} W. D. Hamilton, {\it Narrow Roads of Gene Land: Evolution of Social Behaviour} (Oxford University Press, Oxford, 1996).
\bibitem{Szabo2002} G. Szab\'o and C. Hauert, Phys. Rev. Lett. {\bf 89}, 118101 (2002); C. Hauert and G. Szab\'o,  Am. J. Phys. {\bf 73} 405 (2005).
\bibitem{LV} A. J. Lotka, Proc. Natl. Acad. Sci. U.S.A. {\bf 6}, 410 (1920); V. Volterra, Mem. Accad. Lincei {\bf 2}, 31 (1926).
\bibitem{May74} R. M. May, {\it Stability and Complexity in Model Ecosystems} (Cambridge University Press, Cambridge, England, 1974)
\bibitem{Mobilia-2007} M. Mobilia, I.~T. Georgiev, and U.~C. T\"auber, J. Stat. Phys. {\bf 128}, 447 (2007).
\bibitem{NowakTaylor} M.~A. Nowak, A. Sasaki, C. Taylor, and D. Fudenberg, Nature (London) {\bf 428}, 646 (2004).
\bibitem{PopGen} R. A. Fisher, {\it The Genetical Theory of Natural Selection} (Clarendon, Oxford, 1930); S. Wright, Genetics {\bf 16}, 97 (1931); P. A. P. Moran. {\it The Statistical Processes of Evolutionary Theory} (Clarendon Press, Oxford, 1962).
\bibitem{Nowak2005} M. A. Nowak, Science {\bf 314}, 1560 (2005).
\bibitem{Trivers} R. L. Trivers, Quart. Rev. Biol. {\bf 45} (1971).
\bibitem{Axelrod} R. Axelrod and W.D. Hamilton, Science {\bf 211}, 190 (1981).
\bibitem{Kimura} J.~F. Crow and M. Kimura, {\it An Introduction to Population Genetics Theory} (Harper and Row, New York, 1970); M. Kimura and T. Ohta, {\it Theoretical Aspects of Population Genetics} (Princeton University Press, Princeton, 1971); M. Kimura, {\it The Neutral Theory of Molecular Evolution}
(Cambridge University Press, Cambridge, U.K., 1983).
\bibitem{Hubbell} S. P. Hubbell, {\it The Unified Neutral Theory of Biodiversity and Biogeography} (Princeton University Press, Princeton, 2001).
\bibitem{Bell} G. Bell, Science {\bf 293}, 2413 (2001).
\bibitem{Blythe} R.~A. Blythe and A~J. McKane,  J. Stat. Mech.  {\bf P07018} (2007).
\bibitem{Ohta} T. Ohta, Proc. Natl. Acad. Sci. U.S.A. {\bf 99}, 16134 (2002).
\bibitem{Ohtsuki} H. Ohtsuki {\it et al.}, Nature {\bf 441}, 502 (2006).
\bibitem{Santos} F. C. Santos and J. M. Pacheco, Phys. Rev. Lett. {\bf 95}, 098104 (2005).
\bibitem{Altrock} P.~M. Altrock and A. Traulsen, New J. Phys. {\bf 11}, 013012 (2009).
\bibitem{Antal} T. Antal, I. Scheuring, Bull. Math. Biol. {\bf 68}, 1923 (2006).
\bibitem{Traulsen2006} A. Traulsen, M.~A. Nowak and J.~M. Pacheco, Phys. Rev. E {\bf 74}, 011909 (2006).
\bibitem{noise} H. H. McAdams and A. Arkin, Trends Genet. {\bf 15}, 65 (1999); M. Assaf and B. Meerson, Phys. Rev. Lett. \textbf{100}, 058105 (2008).
\bibitem{Gardiner} C. W. Gardiner, {\it Handbook of Stochastic Methods}, (Springer, New York, 2002), second edition.
\bibitem{vanK} N. G. van Kampen, {\it Stochastic Processes in Physics and Chemistry}, (North-Holland, Amsterdam, 1992).
\bibitem{Ewens} W. J. Ewens, {\it Mathematical Population Genetics} (Springer, New York, 2004), second edition.
\bibitem{Hauert-preprint} A. Traulsen and C. Hauert, arXiv:0811.3538v1
\bibitem{Pacheco} A. Traulsen, J.~M. Pacheco and L.~A. Imhof, Phys. Rev. E {\bf 74}, 021905 (2006).
\bibitem{Traulsen} A. Traulsen, J.~C. Claussen, and C. Hauert, Phys. Rev. Lett. {\bf 95}, 238701 (2005).
\bibitem{o1} One can also account for self-interaction terms of order ${\cal O}(N^{-1})$ in $T^{\pm}(n)$~\cite{AM}. However, while this unnecessarily complicates the mathematical treatment, it does not bring about any qualitative differences in the results.
\bibitem{Bender} L.~D. Landau and E.~M. Lifshitz, {\it Quantum Mechanics: Non-Relativistic Theory} (London, Pergamon, 1977);
C.~M. Bender and S.~A. Orszag, {\it Advanced Mathematical Methods for Scientists and Engineers} (Springer, New York,1999).
\bibitem{kubo} R. Kubo, K. Matsuo, and K. Kitahara, J. Stat. Phys. \textbf{9}, 51 (1973); H. Gang, Phys. Rev. A \textbf{36}, 5782 (1987).
\bibitem{dykman} M.I. Dykman, E. Mori, J. Ross, and P.M. Hunt, J. Chem. Phys. \textbf{100}, 5735 (1994); D.A. Kessler and N.M. Shnerb, J. Stat. Phys. \textbf{127}, 861 (2007); B. Meerson and P.V. Sasorov, Phys. Rev. E \textbf{78}, 060103(R) (2008).
\bibitem{EsK} C. Escudero and A. Kamenev, Phys. Rev. E \textbf{79}, 041149 (2009).
\bibitem{AM} M. Assaf and B. Meerson, Phys. Rev. E \textbf{81}, 021116 (2010).
\bibitem{analogy} The WKB treatment of the master equation via ansatz~(\ref{ansatz}) is analogous to the semi-classical treatment of quantum mechanics in terms of $\hbar$~\cite{Bender}.
\bibitem{fpill} B. Gaveau, M. Moreau, and J. Toth, Lett. Math. Phys. \textbf{37}, 285 (1996); C.R. Doering, K.V. Sargsyan, and L.M. Sander, Multiscale Model. and Simul. \textbf{3}, 283 (2005).
\bibitem{MA} M. Mobilia and M. Assaf, Euro. Phys. Lett. \textbf{91}, 10002 (2010).
\bibitem{Assaf} M. Assaf and B. Meerson, Phys. Rev. E \textbf{74}, 041115 (2006);  Phys. Rev. Lett. \textbf{97}, 200602 (2006); Phys. Rev. E \textbf{75}, 031122 (2007).
\bibitem{Claussen} J.~C. Claussen and A. Traulsen, Phys. Rev. E {\bf 74}, 021905 (2005).
\bibitem{init} When $w\to 0$ (almost neutral dynamics) but $Nw\gg 1$ (to ensure the validity of the WKB approach),  $n$ has to be sufficiently close to $Nx^*\gg 1$ to guarantee the convergence to the coexistence state prior to fixation.
\bibitem{fermi} The case of the FP is degenerate in the sense that, with the special choice of transition rates (\ref{FP}), $\Theta_{\rm WKB}(x)$ is a linear function of $x$ that incidentally coincides with  $\Theta_{\rm   \ell FPE}(x)$. This coincidence, without any general implications, explains why the diffusion approximation~(\ref{fixmp}) with the exponent $\Theta$ given by (\ref{fixfp_l}) works so well for the FP~\cite{Traulsen2006,Hauert-preprint}.
\bibitem{RPS} T. Reichenbach, M. Mobilia, and E. Frey, Nature {\bf 448}, 1046 (2007); Phys. Rev. Lett. {\bf 99}, 238105 (2007); J. Theor. Biol. {\bf 254}, 368 (2008).
\end{thebibliography}
\end{document}